\newcommand{\beq}{\begin{eqnarray}}
\newcommand{\eeq}{\end{eqnarray}}

\newcommand{\dX}{\frac{d{\bf X}}{dt}}

\newcommand{\bfu}{{\bf u}}

\newcommand{\bfU}{{\bf U}}
\newcommand{\bfX}{{\bf X}}
\newcommand{\bfF}{{\bf F}}

\newcommand{\tbfX}{\tilde{\bf X}}
\newcommand{\eps}{\epsilon}

\newcommand{\e}{{\rm e}}

\renewcommand{\d}{\partial}

\renewcommand{\theequation}{\thesection.\arabic{equation}}

\newcommand{\cl}{\centerline}

\newcommand{\btem}{\bibitem}

\newcommand{\PRL}{Phys.\ Rev. \ Lett.}

\documentstyle[12pt]{article}    
\begin{document}
{\Large {\bf Uses of Envelopes for Global and Asymptotic Analysis;\ }}
 {\large{\bf geometrical meaning of renormalization group equation}}
\footnote{Talk presented at RIMS (Kyoto) Workshop on Geometrical Methods 
for Asymptotic Analysis   1997.5.20 -- 5.23}\\ 

\begin{large}
{\cl {Teiji Kunihiro}}
\begin{center}
Faculty of Science and Technology, Ryukoku University,\\ 
Seta, Ohtsu, 520-21, Japan\\ 
\end{center}
\end{large}

\begin{abstract}
We give a comprehensive review of the renormalization group method
 for global and asymptotic analysis, putting an emphasis on the relevance
 to  the classical theory of envelopes and the existence of 
 invariant manifolds of the dynamics under consideration. 
We clarify that an essential point of the method is to  
 convert the problem from solving differential equations to obtaining
 suitable initial (or boundary) conditions. 
 We mention that the notion of envelopes is also useful for
 constructing global and asymptotic behavior of wave functions of 
 quantum systems such as the ones with the quartic potential or
 double-well potential.
\end{abstract}
 
\setcounter{equation}{0}
\renewcommand{\theequation}{\thesection.\arabic{equation}}
\section{Introduction}
The renormalization group (RG) equations 
have a peculiar power to improve the global nature of functions obtained 
in the perturbation theory  in quantum field theory \cite{rg}:
 The RG equations represent  the fact that 
a  physical quantity ${\cal O}(p,\alpha, \mu)$ 
should not depend on the renormalization point $\mu$,
\beq
\frac{\partial {\cal O}(p, \alpha;\mu)}{\d \mu}=0.
\eeq 
Such a floating renormalization
 point was first introduced by Gell-Mann and Low in the celebrated 
paper\cite{rg}. We remark that the renormalization point $\mu$ plays
 a role of the initial point and the  renormalization condition
 for physical quantities such as coupling constants 
at  the energy scale $\mu$ may be viewed as setting initial values of
 these quantities.\cite{shirkov}

Recently, the present author has indicated  that 
the RG equation  $\grave{a}$ la Gell-Mann-Low can be identified 
 as  the envelope equation\cite{kunihiro1};. 
it was shown that the notion of envelopes is 
  useful for improving the perturbative expansions appearing in
 {\em quantum field theory}.
This identification was realized through an examination of 
  the RG method 
 of Chen, Goldenfeld and Oono for global analysis of differential
 equations\cite{cgo}; they applied the RG equation to numerous problems and
 found that the RG equation gives slow dynamics of the system in question.
 The new point of their method is to utilize secular terms which
 usually  appear  
 when perturbation theory is applied to differential equations; this is
 in contrast to all previous methods, which are formulated on the principle
to avoid the appearance of secular terms. 

Their method was reformulated on the 
 basis of the classical theory of envelopes\cite{kunihiro1,kunihiro2}.
It was  demonstrated that owing to the very envelope equation,
the functions constructed from the solutions in the perturbation theory 
 certainly satisfies the differential equation in question uniformly up 
 to the order with which local solutions is 
constructed. It was also shown \cite{kunihiro3} in a most general setting 
that the RG method gives a reduction of 
dynamics  and clarified that there is a correspondence 
between  the RG method and the reductive perturbation
 method\cite{kuramoto}:
 Some interesting examples were also worked out in this method,
 such as  the forced Duffing, the Lotka-Volterra
 and the Lorenz equations; the first example showed that the method is 
 applicable to not only automonous but also {\em non}-automonous equations,
 the second one showed that the method gives phase equations and the
 last one showed  that {\em center manifolds} of the dynamics can be extracted
 in the method. It is noteworthy that the notion of 
 {\em functional self-similarity} 
 (FSS) extracted as the essence of the RG in \cite{shirkov} is only applicable
 to autonomous equations; see \cite{sasa} for an application of the notion of
 FSS for deducing phase equations. 
The above fact suggests that the 
 notion of envelopes  better represents the underlying mathematics of the 
powerfulness of the RG.
 It was recognized also that when the unperturbed equation has {\em neutrally
 stable} solutions, the RG method works well. It implies that the method is
 applicable when the dynamical system under consideration has invariant 
manifolds and useful to extract the manifolds and the dynamics on the 
 manifolds. 
It was shown that global and asymptotic behavior is obtained
 even for  {\em discrete} 
systems by constructing ``envelopes''
 when the system has neutrally stable solution of the unperturbed 
 equations\cite{matu}.

Subsequently the classical theory of envelopes was applied 
 for getting  asymptotic behaviors of wave functions in {\em quantum 
mechanics}\cite{tanaka}. 
This is an optimized perturbation theory in which
 the perturbation theory is combined with a variational method.
 The key ingredient is to
construct an envelope of a set of 
 perturbative wave functions. This leads to a condition
 similar to that obtained from the principle of minimal 
sensitivity\cite{stevenson}. 
 Applications of the method 
 to quantum anharmonic oscillator and the double well potential show that 
  uniformly valid wave functions  with correct asymptotic behavior
 are obtained in the first-order optimized 
 perturbation even for strong couplings.

In this  report, we will give only basic ingredients  of the
 RG method based on our formulation putting an emphasis to the 
 relevance to envelopes; 
 for detailed account of the method and various applications, please
 refer to Ref.'s\cite{kunihiro1,kunihiro2,kunihiro3,matu,tanaka}.
 In this report, we also clarify that the RG method is most lucidly
 formulated by noting that 
 the method  converts
 the problem from solving differential equations to obtaining
 suitable initial (or boundary) conditions as does the usual RG. 
 In this report, we shall not discuss on partial differential equations
 due to the lack of space. Please refer to \cite{nozaki} as well as
 \cite{cgo,kunihiro2,sasa} for this subject.

\setcounter{equation}{0}
\section{A short review of the classical theory of  envelopes}
 
We here give a brief review of the theory of  envelopes. 
Although the theory can be formulated in higher  
dimensions\cite{kunihiro2,kunihiro3}, 
 we consider here envelope curves,
 for simplicity.

Let $\{C_{\tau}\}_{\tau}$ be a family of curves  parametrized by $\tau$ 
in the $x$-$y$  plane; here $C_{\tau}$ is  represented by the 
 equation $F(x, y, \tau)=0$.
 We suppose 
 that $\{C_\tau\}_{\tau}$ has the envelope $E$, which is represented by the
 equation $G(x, y)=0$. 
The problem is to obtain $G(x, y)$ from $F(x, y,\tau)$.

Now let $E$ and a curve $C_{\tau_0}$ have the common tangent line at 
 $(x,y)=(x_0,y_0)$, i.e., $(x_0, y_0)$ is the point of tangency.
  Then $x_0$ and $y_0$ are functions of $\tau_0$;
$x_0=\phi(\tau_0),\ y_0= \psi(\tau_0)$, and of course $ G(x_0, y_0)=0$. 
Conversely, for each point $(x_0, y_0)$ on $E$, there exists a  parameter
 $\tau_0$.
 So we can reduce the problem to get $\tau_0$ as a function of 
$(x_0,y_0)$;
 then $G(x, y)$ is obtained as 
$F(x, y, \tau(x,y))=G(x,y)$.\footnote{Since there
 is a relation $G(x_0, y_0)=0$ between $x_0$ and $y_0$, $\tau_0$ is actually a
 function of $x_0$ {\em or } $y_0$.} $\tau_0(x_0, y_0)$ can be obtained as 
 follows.

Since the tangent line of $E$ at $(x_0, y_0)$  is perpendicular to the normal
 direction of $F(x, y, \tau)=0$ at the same point,  one has
\beq
F_x(x_0, y_0, \tau_0)\phi'(\tau_0)+ F_y(x_0, y_0, \tau_0)\psi'(\tau_0)=0.
\eeq
 On the other hand, differentiating 
$F(x(\tau_0), y(\tau_0), \tau_0)=0$ with respect to $\tau_0$, one also has
\beq
F_x(x_0, y_0, \tau_0)\phi'(\tau_0)+ F_y(x_0, y_0, \tau_0)\psi'(\tau_0)
 + F_{\tau_0}(x_0, y_0, \tau_0)=0.
\eeq
Combining the last two equations, we have
\beq
 F_{\tau_0}(x_0, y_0, \tau_0)
\equiv\frac{\partial F(x_0, y_0, \tau_0)}{\partial \tau_0}=0.
\eeq
This is the basic equation of the theory of envelopes; we call
 this type of equation envelope equations and also RG/E equation where
 RG and E stand for renormalization group and envelope, respectively,
 becasue the RG equation Eq(1.1) has the same form as Eq.(2.3).
One can thus 
 eliminate the parameter $\tau_0$ to get a relation between  $x_0$ and $y_0$,
\beq
G(x, y) = F(x, y, \tau_0(x, y))=0,
\eeq
 with the replacement $(x_0, y_0)\rightarrow (x,y)$. $G(x, y)$ is called
 the discriminant of $F(x, y, t)$. 

Comments are in order here: 
(i)\  When the family of curves is given by the
 function $ y=f(x, \tau)$, the condition Eq.(2.3) is reduced to 
${\partial f}/{\partial \tau_0}=0$; the envelope is given by 
 $y=f(x, \tau_0(x))$. 
(ii)\  The equation $G(x, y)=0$ may give 
 not only the envelope $E$ but also a set of 
singularities of the curves $\{C_{\tau}\}_{\tau}$. This is because 
 the condition that 
$\partial F/\partial x = \partial F/\partial y=0$ is also 
 compatible with  Eq. (2.3).

\setcounter{equation}{0} 
\section{The RG method; a simplest example}
\renewcommand{\theequation}{\thesection.\arabic{equation}}

In this section, using a simplest example we show how the RG 
method  works for 
 obtaining  global and asymptotic behavior of solutions of differential
 equations. We shall present the method so that the reader 
will readily see that the 
notion of envelopes is 
intrinsically related to the method.
We shall emphasize that an essential point of the method is tuning the
 initial condition  at an arbitrary time $t_0$ perturbatively along with 
 solving the perturbative  equations successively. 
One will see that the reasoning for various steps in the procedure
 and the underlying picture are quite different from the
 original ones given in \cite{cgo}.
We believe, however, that the  present formulation  emphasizing the role of
initial conditions 
 and the relevance to envelopes of perturbative local solutions 
straightens the
 original argument, and is  the most 
comprehensive one.\footnote{The following is even a refinment 
of the argument given in \cite{kunihiro1,kunihiro3} where
 the fact that the RG method is  a theory manipulating  initial conditions
 were not fully recognized.}

Let us take the following simplest example to show 
our method:
\beq
\frac{d^2 x}{dt^2}\ +\ \eps \frac{dx}{dt}\ +\ x\ =\ 0,
\eeq
where $\eps$ is supposed to be small. The solution to Eq.(3.1) reads
\beq
 x(t)= \bar {A} \exp (-\frac{\eps}{2} t)\sin( \sqrt{1-\frac{\eps^2}{4}} t 
+ \bar {\theta}),
\eeq
where $\bar {A}$ and $\bar{\theta}$ are constants.

 Now, let us  obtain the solution around the initial time $t=t_0$ in a
 perturbative way,  expanding $x$ as
\beq
x(t, t_0) = x_0(t, t_0) \ +\ \eps  x_1(t ,t_0)\ +\ \eps ^2 x_2(t, t_0)\ 
+\ ... ,
\eeq
 where $x_n(t, t_0)$ ($n= 0, 1, 2 ...$) satisfy
\beq
\ddot{x}_0  +\ x_0\ =\ 0,
\ \ \ \
 \ddot{x}_{n+1}\ +\ x_{n+1}\ =- \dot{x}_n . 
\eeq
The initial condition may be specified by
\beq
x(t_0, t_0)= W(t_0).
\eeq
We suppose that the initial value $W(t_0)$ is always on an exact solution of
 Eq.(3.1) for any  $t_0$. We also expand the initial value $W(t_0)$;
\beq
W(t_0) = W_0(t_0) \ +\ \eps  W_1(t_0)\ +\ \eps ^2 W_2(t_0)\ 
+\ ... ,
\eeq
and the terms $W_i(t_0)$ will be determined so that the perturbative solutions
 around different initial times $t_0$  have an 
 envelope.  Hence the initial value $W(t)$ thus constructed will give 
the (approximate but) global solution of the equation. 

Let us perform the above program. 
The lowest solution may be given by
\beq
x_0(t, t_0) = A(t_0)\sin (t +\theta (t_0)), 
\eeq
where we have made it explicit that the constants $A$ and $\theta$ may 
depend on the initial time $t_0$.
The initial value $W(t_0)$ as a function of $t_0$
 is specified as
\beq
W_0(t_0)= x_0(t_0,t_0)= A(t_0)\sin (t_0 +\theta (t_0)).
\eeq
We remark that Eq.(3.7) is a neutrally stable solution; with 
 the perturbation $\eps \not=0$ 
 the constants $A$ and $\theta$ may move slowly.  We shall see that 
 the envelope equation gives the equations describing the slow motion of
 $A$ and $\theta$.
 
The first order equation now reads
$\ddot{x}_1\ +\ x_1\ =- A\cos(t+\theta),$
 and we choose the solution in the following form,
\beq
x_1(t, t_0)= -\frac{A}{2}\cdot (t -t_0)\sin(t+\theta),
\eeq 
which means that the first order initial value $W_1(t_0)=0$ so that
the lowest order value $W_0(t_0)$ approximates the exact value as closely 
 as possible.
Similarly, the second order solution may be given by
\beq
 x_2(t)= 
\frac{A}{8}\{ (t-t_0)^2\sin(t +\theta) - (t-t_0)\cos(t+\theta)\},
\eeq
 thus $W_2(t_0)=0$ again for the present linear equation.

It should be noted  that  the secular terms have appeared 
 in the higher order terms, which are  absent in the 
exact solution and invalidates the perturbation theory for $t$ far
  from $t_0$. However, with the very existence of the secular terms,
 we could make $W_i(t_0)$ ($i=1, 2$) vanish and  $W(t_0)=W_0(t_0)$
 up to the third order.

Collecting the terms, we have 
\beq
x(t, t_0)&=& A\sin (t +\theta) -\eps\frac{A}{2} (t -t_0)\sin(t+\theta)
  \nonumber \\ 
 \ \ \ & \ \ \ & +\eps^2\frac{A}{8}
\{ (t-t_0)^2\sin(t +\theta) - (t-t_0)\cos(t+\theta)\},
\eeq
and more importantly
\beq
W(t_0)=W_0(t_0)=A(t_0)\sin (t_0 +\theta (t_0))
\eeq
 up to $O(\eps^3)$. We remark that 
$W(t_0)$ describing the solution 
 is parametrized by possibly slowly moving variable $A(t_0)$ and 
 $\phi (t_0)\equiv t_0+\theta (t_0)$ in a definite way.

Now we have a family of curves $\{C_{t_0}\}_{t_0}$ given by functions 
$\{x(t, t_0)\}_{t_0}$ parametrized with $t_0$. 
 They are all on the exact curve $W(t)$ at $t=t_0$ 
 up to $O(\eps ^3)$, but 
 only valid locally for $t$ near $t_0$. 
 So it is conceivable that the envelope 
 $E$ of $\{C_{t_0}\}_{t_0}$ which 
 contacts with each local solution at $t=t_0$ will give a global solution.
 Thus the envelope function $x_{_E}(t)$ coincides with $W(t)$;  
\beq
x_{_E}(t)=x(t,t)=W(t).
\eeq  
Our task is actually to determine $A(t_0)$ and $\theta(t_0)$ as 
 functions of $t_0$ so that the family of the local solutions has an 
 envelope.
 According to the classical theory of envelopes given in the previous
 section, the above program can be achieved by
 imposing that the envelope equation 
\beq
 \frac{dx(t, t_0)}{d t_0}=0,
\eeq
 gives the solution $t_0=t$.
From Eq.'s (3.11)  and (3.14), we have
\beq
\frac{dA}{dt_0} + \eps A =0,  \ \ \ 
\frac{d\theta}{dt_0}+\frac{\eps^2}{8}=0,
\eeq
where we have used the fact that $dA/dt$ is $O(\eps)$ and neglected
 the terms of $O(\eps^3)$.
Solving the equations, we have 
\beq
A(t_0)= \bar{A}{\rm e}^{-\eps t_0/2}, \ \ \ 
\theta (t_0)= -\frac{\eps^2}{8}t_0 + \bar{\theta},
\eeq
where $\bar{A}$ and $\bar{\theta}$ are constant numbers.
 Thus we get
\beq
x_{_E}(t)= x(t, t)=W_0(t)= 
\bar{A}\exp(-\frac{\eps}{2} t)\sin((1-\frac{\eps ^2}{8})t + 
\bar{\theta}),
\eeq
up to $O(\eps^3)$.
Noting that $\sqrt{1 - {\eps^2}/{4}}= 1 - {\eps^2}/{8} + O(\eps ^4)$, 
 one finds
 that the resultant envelope function $x_{_E}(t)=W_0(t)$ is an approximate but 
{\em  global} solution to Eq.(3.1); see Eq. (3.2).

\setcounter{equation}{0}
\section{Nonlinear equations}
\setcounter{equation}{0}
\renewcommand{\theequation}{\thesection.\arabic{equation}}

In this section, we treat a couple of examples of systems
of ODE's with nonlinearity to show how the RG method works\cite{kunihiro3}. 
 The examples are  the  Lotka-Volterra\cite{lotka} and the Lorenz\cite{lorenz} 
equation.  
We shall  derive the time dependence  of the
 solution to the Lotka-Volterra equation explicitly; a phase equation will 
 be derived by our method.
 The Lorenz equation  is an example with three degrees of freedom, which shows 
a  bifurcation.  We shall give the center manifolds of this
 equation around the first bifurcation point.

\subsection{Lotka-Volterra equation}
     
The Lotka-Volterra equation reads\cite{lotka};
\beq
\dot{x}= ax -\eps xy, \ \ \ \ \dot{y}=-by+\eps'xy,
\eeq
where the constants $a, b, \eps$ and $\eps'$ are assumed to be positive.
It is well known that the equation has the  conserved quantity, i.e., 
\beq
b\ln\vert x\vert + a\ln \vert y\vert -(\eps' x+\eps y)={\rm const.}.
\eeq 

The fixed points are given by $(x=0, y=0)$ and $(x=b/\eps', y=a/\eps)$.
Shifting and scaling  the variables by
\beq
x=(b+ \eps\xi)/\eps', \ \ \ \ y=a/\eps + \eta,
\eeq
we get the reduced equation given by the system
\beq
\biggl(\frac{d}{dt}- L_0\biggl)\bfu= -\eps\xi\eta\pmatrix{\ 1\cr -1},
\ \ \ \ 
\eeq
where  
\beq
\bfu = \pmatrix{\xi\cr \eta},\ \ \ \ L_0=\pmatrix{0 & -b\cr a & \ 0}.
\eeq
The eigen value equation
\beq
L_0\bfU=\lambda _0\bfU
\eeq
has the solution
\beq
\lambda _0=\pm i\sqrt{ab}\equiv \pm i\omega, \ \ \ \ 
\bfU =\pmatrix{\, 1\cr \mp i\frac{\omega}{b}}.
\eeq

Let us try to extract the global behavior of the solution around the
 fixed point.  Our strategy is the following: We suppose that we are
 on the exact solution at $t=t_0$ where $t_0$ is arbitrary; we denote
 the initial value by ${\bf W}(t_0)$.
 We also suppose that we can apply perturbation theory for the solution
  at least in the small neighborhood of $t=t_0$.
We   expand the variable in a Taylor series of $\eps$;
\beq
\bfu=\bfu_0+\eps\bfu_1 +\eps^2\bfu_2+\cdots,
\eeq
with $\bfu _i=\ ^t(\xi _i, \eta_i)$. An essential point of our method
 is to expand the initial value, too;
\beq
{\bf W}(t_0)={\bf W}_0(t_0)+\eps{\bf W}_1(t_0) +
\eps^2{\bf W}_2(t_0)+\cdots .
\eeq
Our central task is to extract the initial value as 
 a function of $t_0$ so that the resulting local solutions starting
 from the different initial points at $t=t_0$ and , say, $t=t_0+\Delta t$
 are continued smoothly.  This  condition is found to be nothing but the
 one that the local solutions have an envelope.
In actual calculations, it is also important to use the fact that 
the functional form of the initial values  can be reduced from the
 general solution of the differential equations in a perturbative way;
 in this procedure, only independent functions modulo to secular terms
 are retained.

  The lowest term satisfies the equation
\beq
\biggl(\frac{d}{dt}- L_0\biggl){\bfu}_0={\bf 0},
\eeq
which yields the solution
\beq
\bfu _0(t;t_0)={\cal A}(t_0){\e}^{i\omega t}\bfU + {\rm c.c.}.
\eeq
 Notice that ${\cal A}$ is a complex number, so one may parametrize it as
 \beq
{\cal A}(t_0)=A(t_0)/2i\cdot{\rm exp}(i \theta(t_0)).
\eeq
 The solution implies that the initial condition is given by
\beq 
\bfu _0(t_0; t_0)={\cal A}(t_0){\e}^{i\omega t_0}\bfU + {\rm c.c.}.
\eeq
It means that in the lowest approximation the solution is
 parametrized by a complex function ${\cal A}(t_0)$ or a pair of real 
functions, $A(t_0)$ and $\phi(t_0)\equiv \omega t_0+ \theta(t_0)$.
With a small perturbation, we expect that $A$ and $\theta$ will move
 slowly.

Noting that 
$\pmatrix{\ 1\cr -1}=\alpha \bfU + {\rm c.c.},$
with $\alpha=(1- ib/\omega)/2$, one finds that 
the first order term  satisfies the equation
\beq
\biggl(\frac{d}{dt} - L_0\biggl)\bfu _1=
 \frac{\omega}{b}\biggl[i{\cal A}^2 \e ^{2i\omega t}
  (\alpha \bfU + {\rm c.c.}) + {\rm c. c.}\biggl],
\eeq
  the solution to which may be given by
\beq
\bfu _1=\frac{1}{b}\biggl[{\cal A}^2(\alpha \bfU + \frac{\alpha ^{\ast}}{3}
\bfU ^{\ast})
 \e^{2i\omega t} + {\rm c.c.}\biggl].
\eeq
Thus the initial value ${\bf W}_1(t_0)$ in this order is given
$\bfu _1(t_0, t_0)$.

Similarly, the second  order solution may be given by 
\beq
\bfu _2&=& \Biggl[\frac{b-i\omega}{3b^2}\vert {\cal A}\vert ^2{\cal A}
\biggl\{ \alpha (t-t_0 +i\frac{\alpha^{\ast}}{2\omega})
         \bfU + \frac{\alpha ^{\ast}}{2i\omega}\bfU ^{\ast}\biggl\}
           \e ^{i\omega t } \nonumber \\ 
 \ \ \  & & + \frac{b+i\omega}{4b^2i\omega}{\cal A}^3(2\alpha \bfU +
 \alpha^{\ast}\bfU ^{\ast})\e^{3i\omega t}\Biggl]
+ {\rm c.c.} .
\eeq
Here we have a secular term proportional to the unperturbed solution.
Since we want to make the lowest initial value as close as the
 exact one, we demand that as many as possible terms in the higher order
 vanish at $t=t_0$. Thus adding unperturbative solutions,
 we make the secular term (of the upper component) vanishes at 
$t=t_0$.\footnote{Although we can make other terms also vanish at 
$t=t_0$\cite{devega}, the resulting dynamics of $A$ and $\theta$ become
 more complicated than the present choice. In the theory of reduction of 
dynamics, one usually prefers simpler dynamics\cite{kuramoto2}.}
Thus neglecting  higher order terms, 
we have  $\bfu (t, t_0) =\bfu _0 +\eps \bfu _1 + \eps ^2 \bfu _2$, and 
 ${\bf W}(t_0)= \bfu (t_0, t_0)$.   We impose that the solutions
  $\bfu (t, t_0)$ and  $\bfu (t, t_0+\Delta t)$\ give
 the same value at $t$. By taking a limit $\Delta t \rightarrow 0$,
 we have  the envelope equation; 
\beq
\frac{d \bfu}{d t_0}={\bf 0},
\eeq
 with $t_0=t$.  This gives the equation for ${\cal A}(t)$ as
\beq
\frac{d {\cal A}}{dt}= - i\eps^2 \frac{\omega ^2+b^2}{6\omega b^2}\vert {\cal A}\vert ^2 {\cal A}.
\eeq 
In terms of  $A(t)$ and $\theta (t)$, we have
\beq
A(t)= {\rm const.}, \ \ \ \ 
\theta (t) = - \frac{\eps^2A^2}{24}(1+ \frac{b^2}{\omega ^2})\omega t
  + \bar{\theta }_0,
\eeq
with $\bar{\theta }_0$ being a constant.  Owing to the  prefactor $i$ 
in r.h.s. of Eq. (4.18), the absolute value of the amplitude $A$ becomes
 independent of $t$, while the phase $\theta$ has a $t$-dependence.
  The envelope function is given by
\beq
\bfu _E(t)=\pmatrix{\xi _E(t)\cr \eta _E(t)}=
\bfu (t, t)={\bf W}(t).
\eeq
In terms of the components, one has
\beq
\xi _{_E}&= & A\sin \Theta (t) -
\eps \frac{A^2}{6\omega}(\sin 2\Theta (t)
 + \frac{2\omega }{b}\cos 2\Theta (t))\nonumber \\ 
 \ \ \ & \ & -\frac{\eps^2 A^3}{32}\frac{3\omega ^2 -b^2}{\omega ^2b^2}
(\sin 3\Theta (t) - \frac{4\omega b}{3\omega ^2 -b^2}\cos 3\Theta (t) ),
\nonumber \\ 
\eta _{_E} &=& -\frac{\omega}{b}\Biggl[
 \biggl(A - \frac{\eps^2A^3}{24}\frac{b^2-\omega ^2}{b^2\omega ^2}\biggl)
\cos \Theta (t) - \frac{\eps ^2 A^3}{12b\omega}\sin \Theta (t)
 \\ 
 \ \ \ \ & \ & + \eps \frac{A^2}{2b}\biggl(\sin 2\Theta (t) - 
\frac{2b}{3\omega}\cos 2\Theta (t)\biggl)
  - \frac{\eps^2A^3}{8b\omega}\biggl( \sin 3\Theta (t)
- \frac{3b^2 -\omega ^2}{4b^2\omega ^2}\cos 3\Theta (t)\biggl)\Biggl],
\nonumber 
\eeq
where
\beq
\Theta (t) \equiv \tilde {\omega} t + \bar{\theta}_0, 
\ \ \ \ \tilde {\omega} \equiv \{ 
1- \frac{\eps^2A^2}{24}(1+ \frac{b^2}{\omega ^2})\}\omega .
\eeq
One sees that the angular frequency is shifted.

 We can identify $\bfu_E(t)= (\xi _E(t), \eta _E(t))={\bf W}(t)$ 
as an approximate 
 solution to Eq.(4.4) by construction. 
We see that 
  $\bfu_E(t)$
is an approximate but uniformly valid solution to the equation
 up to $O(\eps^3)$.  We remark that 
 the resultant trajectory is closed in conformity  with the conservation law 
Eq. (4.2).

``Explicit solutions'' of two-pieces of Lotka-Volterra equation were
 considered by Frame \cite{frame}; however, his main concern was on 
 extracting the  period of the solutions in an average method.
 we are  not aware of any other 
work than ours 
which gives an explicit form of the solution as given here.

\subsection{The Lorenz model}

The Lorenz model\cite{lorenz} for the thermal convection is given by 
\beq
\dot{\xi}&=&\sigma(-\xi+\eta),\nonumber \\
\dot{\eta}&=& r\xi -\eta -\xi\zeta,\nonumber \\
\dot{\zeta}&=& \xi\eta - b \zeta.
\eeq
The steady states are give by 
\beq
{\rm (A)}\ \ (\xi, \eta, \zeta)=(0, 0, 0),\ \ \ 
{\rm (B)}\ \ (\xi, \eta, \zeta)=
(\pm \sqrt{b(r-1)},\pm \sqrt{b(r-1)},r-1).
\eeq

The linear stability analysis\cite{holmes} shows that the origin is stable for 
$0<r<1$ but unstable for $r>1$, while the latter steady states (B) are 
 stable for $1<r<\sigma(\sigma+b+3)/(\sigma -b-1)\equiv r_c$ but 
unstable for $r>r_c$.
In this report, we examine the non-linear stability around the origin
 for $r\sim 1$ and extract a center manifold for the dynamics in this 
 region.

We put 
\beq
r=1+\mu \ \ \ {\rm and}\ \ \ 
\mu =\chi \eps^2, \ \ \ \chi={\rm sgn}\mu.
\eeq
We expand the quantities as Taylor series of $\eps$:
\beq
\bfu\equiv \pmatrix{\xi\cr 
                   \eta\cr 
                   \zeta}
 = \eps \bfu_1+\eps^2\bfu_2 + \eps ^3\bfu_3 + \cdots,
\eeq
where $\bfu _i=\ ^t(\xi_i, \eta_i, \zeta_i) $\ $(i=1, 2, 3, \dots)$.
 We also expand the initial value at $t=t_0$;
\beq
\bfu (t_0, t_0)= {\bf W}(t_0)
 = \eps {\bf W}_1+\eps^2{\bf W}_2 + \eps ^3{\bf W}_3 + \cdots.
\eeq

The first order equation reads
\beq
\biggl(\frac{d}{dt} - L_0\biggl)\bfu_1={\bf 0},
\eeq
where
\beq
L_0=\pmatrix{-\sigma & \sigma & 0\cr 
                 1   &    -1  & 0\cr
                 0   &    0   & -b},
\eeq
the eigenvalues of which are found to be
\beq
\lambda _1=0, \ \ \ \lambda _2= - \sigma -1,\ \ \ \lambda _3= -b.
\eeq
The respective eigenvectors are 
\beq
\bfU _1=\pmatrix{1\cr
                 1\cr
                 0}, \ \ \ 
\bfU _2=\pmatrix{\sigma\cr
                 -1\cr
                 0}, \ \ \ 
\bfU _3=\pmatrix{0\cr
                 0\cr
                 1}.
\eeq

We are interested in the {\em asymptotic state} as $t\rightarrow \infty$.
In this asymptotic region, one may take only the neutrally stable 
solution
\beq
\bfu _1(t; t_0)=A(t_0)\bfU_1,
\eeq
because the other terms proportional to $\bfU _{2, 3}$ will decay out
 at a sufficiently large time.
Here we have made it explicit that the solution may depend on the
 initial time $t_0$. Eq.(4.32) implies that we have taken the initial 
condition that 
\beq
{\bf W}(t_0)\simeq {\bf W}_1(t_0)=A(t_0)\bfU_1.
\eeq
  In terms of the components,
\beq
\xi_1(t)=A(t_0), \ \ \ \eta_1(t)=A(t_0), \ \ \ \zeta _1(t) =0.
\eeq
In another word, the motion of $\bfu _1(t)$ is confined or reduced 
to the one-dimensional manifold $^t (A, A, 0)$, 
although  $A$ is  a constant in this approximation.
 One expects that the small perturbation with $\eps\not=0$ will 
 give rise to a slow motion of $A$ as well as a modification of the 
 slow manifold.

The second order equation  reads
\beq
\biggl(\frac{d}{dt} - L_0\biggl)\bfu_2=\pmatrix{\ \ 0\cr
                                                -\xi_1\zeta_1\cr
                                                \xi_1\eta_1}
                                     = A^2\bfU_3,
\eeq
which may yield
\beq
\bfu_2(t)=\frac{A^2}{b}\bfU_3,
\eeq
 or in terms of the components
\beq
\xi_2=\eta_2=0, \ \ \  \zeta_2=\frac{A^2}{b}.
\eeq
Here we have retained only functions independent of the ones appearing
 in the lowest approximation. 
Then the third order equation is given by
\beq
\biggl(\frac{d}{dt} - L_0\biggl)\bfu_3= 
      \pmatrix{\ \ \ 0\cr
               -\chi\xi_1-\xi_2\zeta_1-\xi_1\zeta_2\cr
               \xi_2\eta_1+\xi_1\eta_2}
    = \frac{1}{1+\sigma}(\chi A-\frac{1}{b}A^3)(\sigma\bfU_1 -\bfU_2),
\eeq
which may yield
\beq
\bfu_3=\frac{1}{1+\sigma}(\chi A-\frac{1}{b}A^3)
      \{\sigma(t-t_0 + \frac{1}{1+\sigma})\bfU_1 -
   \frac{1}{1+\sigma}\bfU_2\}.
\eeq
Here we have again retained functions which have not appeared before
 except for the term with  which the secular terms vanishes at $t=t_0$.
  Of course, one may have other
 choices for the independent functions, but it is found that the 
present ``minimal'' choice gives the simplest dynamics for the amplitude
 $A$.

Thus collecting all the terms, one has 
\beq
\bfu (t;t_0)&=& 
\eps A(t_0)\bfU_1 + \frac{\eps^2}{b}A(t_0)^2\bfU_3 \nonumber \\ 
 \ \ \ \ &\ & \ \ \ 
 + \frac{\eps ^3}{1+\sigma}(\chi A(t_0) -\frac{1}{b}A(t_0)^3)
      \{\sigma(t-t_0 + \frac{1}{1+\sigma})\bfU_1 -
   \frac{1}{1+\sigma}\bfU_2\},
\eeq
 up to $O(\eps ^4)$. Accordingly, the initial value reads
\beq
{\bf W}(t_0)= \bfu (t_0, t_0)&=& 
\eps A(t_0)\bfU_1 + \frac{\eps^2}{b}A(t_0)^2\bfU_3 \nonumber \\ 
 \ \ \ \ &\ & \ \ \ 
 + \frac{\eps ^3}{(1+\sigma)^2}(\chi A(t_0) -\frac{1}{b}A(t_0)^3)
      \{\sigma\bfU_1 - \bfU_2\}.
\eeq

Demanding that the solutions at different initial times
 are continued smoothly, we have the  RG/E equation, which reads
\beq
{\bf 0}&=&\frac{d \bfu}{d t_0}\biggl\vert_{t_0=t},\nonumber \\ 
\ &=& \eps \frac{dA}{dt}\bfU_1+ 2 \frac{\eps^2}{b}A\frac{dA}{dt}\bfU_3
 -\frac{\sigma}{1+\sigma}\eps^3(\chi A - \frac{1}{b}A^3)\bfU_1,
\eeq 
up to $O(\eps^4)$. Noting that one may self-consistently assume that 
 $dA/dt=O(\eps^2)$,  we have the amplitude equation
\beq
\frac{dA}{dt}=\eps^2\frac{\sigma}{1+\sigma}(\chi A(t) - \frac{1}{b}A(t)^3).
\eeq
With this $A(t)$, the envelope function is given by the initial value 
\beq
\bfu_E(t)&=&\bfu (t; t_0=t), \nonumber \\ 
 \ \  &=& {\bf W}(t),\nonumber \\ 
 \ \ \ &=& 
\eps A(t)\bfU_1 + \frac{\eps^2}{b}A(t)^2\bfU_3 
+ \frac{\eps ^3}{(1+\sigma)^2}(\chi A(t) -\frac{1}{b}A(t)^3)
      (\sigma \bfU_1 -\bfU_2),
\eeq
or
\beq
\xi_E(t)=\eps A(t),\ \ 
\eta_E(t)= \eps A(t) +\frac{\eps^3}{1+\sigma}
           (\chi A(t)-\frac{1}{b}A(t)^3),\ \ 
\zeta_E(t)= \frac{\eps^2}{b}A(t)^2.
\eeq
We see that the initial value is obtained as the envelope of the
 local solutions and becomes
  a global
 solution to the Lorenz model.

A remark is in order here; Eq.(4.45) shows that the slow manifold which 
 may be identified with a center manifold\cite{holmes} is given by 
\beq
\eta=(1+ \eps^2\frac{\chi}{1+\sigma})\xi - \frac{1}{b(1+\sigma)}\xi^3,
 \ \ \ \zeta= \frac{1}{b}\xi^2.
\eeq
Notice that the invariant manifold is modified with the perturbation and
 also the slow dynamics Eq.(4.43) on the manifold is obtained by our method.
One thus sees that the RG method is a powerful tool to extract center 
manifolds in a concrete form. It is worth mentioning that since the 
RG method utilizes
 neutrally stable solutions as the unperturbed ones, it is rather natural
 that the RG method can extract center manifolds when exist\cite{kuramoto}.

\setcounter{equation}{0}
\section{The basis of the RG method for systems }
\renewcommand{\theequation}{\thesection.\arabic{equation}}

In this section, we give an account of our method in a general
 setting for ordinary equations as a summary.

Let ${\bf X}=\, ^t(X_1, X_2, \cdots , X_n)$ and 
 ${\bf F}({\bf X}, t; \eps) =\, ^t(F _1(\bfX, t; \eps)$, 
$F _2(\bfX, t; \eps),\cdots , F _n(\bfX, t; \eps))$, 
and $\bfX$ satisfy the equation
\beq
\dX = \bfF(\bfX , t; \eps),
\eeq
 with the initial condition ${\bf X}(t_0)= {\bf W}$, where $t_0$ is arbitrary.
 We remark that the initial value ${\bf W}$ may be  dependent
 on $t_0$, i.e., ${\bf W}={\bf W}(t_0)$. We also write the solution
 ${\bf X}$ as ${\bf X}(t; t_0)$ so that the initial-time
dependence is explicit. 
 
Let us try  to have the perturbation solution of 
Eq.(5.1) around $t=t_0$ by expanding 
\beq
\bfX (t; t_0)= \bfX _0(t; t_0) + \eps \bfX _1(t; t_0) 
               + \eps^2\bfX _2(t; t_0) \cdots.
\eeq
We also expand the initial value as
\beq
{\bf X}(t_0, t_0)= {\bf W}(t_0)= {\bf W}_0(t_0) + 
\eps {\bf W}_1(t_0)) + \eps^2 {\bf W}_2(t_0) + \cdots.
\eeq
In fact,  ${\bf X}_i(t_0, t_0)= {\bf W}_i(t_0)$.
We suppose that an approximate solution 
$\tilde{\bfX}=\tbfX (t; t_0)$ 
to the equation up to  $O(\eps ^p)$ is obtained.
It implies that the initial value ${\bf W}(t_0)$ at $t=t_0$
 coincides with an exact solution up to $O(\eps ^p)$.
We also have  
\beq
\frac{d\tbfX  (t; t_0, \tilde{\bf W}(t_0))}{dt}=
\bfF (\tbfX (t; t_0,\tilde{\bf W}(t_0) ), t; \eps) + O(\eps^p).
\eeq

 One may say that now we have a family of the orbits
given  by the functions $\tbfX(t; t_0, \tilde{\bf W}(t_0))$
 with $t_0$  parameterizing the
orbits.
 We see that  the envelope E of  the family of the orbits 
  which contacts with each curve at $t=t_0$ will give an approximate
 but global solution of the equation.
  Thus the envelope function is nothing but the initial value as a function of
 the initial time 
\beq
\bfX _E(t)=\tbfX ( t; t, {\bf W}(t))= \tilde{\bf W}(t).
\eeq
  The construction of E is performed as follows: We impose that the
 RG/E equation
\beq
\frac{d\tbfX}{d t_0}={\bf 0}
\eeq
gives the solution $t_0=t$, from which the dynamics of the 
 initial value ${\bf W}(t)$ is obtained. 
Eq.(5.6) may  give equations as many as $n$ which are  independent 
 of each other. In the applications to describe asymptotic behavior
 of solutions, the equation
 is usually reduced to a low-dimensional equation.

In accord with the above relation, one can easily show that
 $\bfX_E(t)=\tilde{\bf W}(t)$ satisfies the original
 equation uniformly up to $O(\eps ^p)$.
In fact,
 $\forall t_0$,  one has 
\beq
\frac{d\bfX_E}{dt}\Biggl\vert _{t=t_0} &=& 
\frac{d\tbfX(t; t_0, {\bf W}(t_0))}{d t}\Biggl\vert _{t=t_0}+
\frac{d\tbfX(t; t_0, {\bf W}(t_0))}{d t_0}\Biggl\vert _{t=t_0},
 \nonumber \\ 
 \ \ &=& \frac{d\tbfX(t; t_0, {\bf W}(t_0))}{d t}\Biggl\vert _{t=t_0},
\nonumber \\ 
 \ \ &=& \bfF (\bfX _E(t_0), t_0; \eps) + O(\eps^p),
\eeq
 where Eq.(5.6) has been used in the last equality.

\section{Concluding remarks}

We have described the perturbative RG method for global and asymptotic 
analysis. We have emphasized its relevance to the classical
 theory of envelopes and that the method concerns with the initial or boundary
 values of differential equations. 
  It should be remarked  that the notion of envelopes is also 
useful for improvement
 of perturbation series in quantum filed theory\cite{kunihiro1} and
  for obtaining asymptotic behavior of wave functions in quantum 
mechanics \cite{tanaka}.

%\newpage 
\newcommand{\NG}{N. \ Goldenfeld}
\newcommand{\YO}{Y.\ Oono}

\end{document}